# A novel Kagome uud-ddu spin order in Heisenberg spin-1/2 Clinoatacamite $Cu_4(OH)_6Cl_2$, the parent compound of Herbertsmithite

Xu-Guang Zheng[*,1,5], Masato Hagihala[§,2], Ichihiro Yamauchi[1], Eiji Nishibori[3], Takashi Honda[4], Takahiro Yuasa[‡,1], Chao-Nan Xu[5]

[1] Department of Physics, Faculty of Science and Engineering, Saga University, Saga 840-8502, Japan.
[2] Japan Atomic Energy Agency, Ibaraki 319-1184, Japan.
[3] Department of Physics, Faculty of Pure and Applied Sciences and Tsukuba, Research Center for Energy Materials Science, University of Tsukuba, Ibaraki 305-8571, Japan.
[4] Institute of Materials Structure Science, High Energy Accelerator Research Organization, Ibaraki 319-1106, Japan.
[5] Department of Materials Science and Engineering, Graduate School of Engineering, Tohoku University, Sendai 980-8579, Japan.

[*]Corresponding author, E-mail: zheng@cc.saga-u.ac.jp
[§]co-first author.
[‡] Present Address: Institute of Materials Structure Science, High Energy Accelerator Research Organization, Ibaraki 319-1106, Japan.

**Abstract**: The newly identified field-induced up-up-down order in $Ba_3CoSb_2O_9$ etc. renewed attention on exotic phases in spin-1/2 triangular-lattice antiferromagnets. Here, we report a unique zero-field noncoplanar up-up-down—down-down-up Kagome spin order in spin-1/2 antiferromagnet Clinoatacamite, $Cu_4(OH)_6Cl_2$, which consists of weakly-coupled Kagome layers and was known as the parent compound for the most researched spin liquid candidate Herbertsmithite $ZnCu_3(OH)_6Cl_2$. The two-dimensional uud-ddu Kagome order develops below $T_{N1}$=18.1 K in Clinoatacamite before a further transition into a three-dimensional magnetic order at low temperatures below $T_{N2} \sim 6.4$ K with persistent spin fluctuations. The present work reveals a new unpredicted Kagome order in a readily accessible temperature range in the parent compound of a well-studied spin liquid. In addition, it has also solved a puzzling issue for a mysterious magnetic phase.

Quantum spin liquid (QSL) due to geometric frustration in triangular, Kagome and pyrochlore lattices etc. is a central topic in condensed matter physics since Anderson proposed the concept and developed the theory of resonating valence bonds (RVB) to account for high-temperature superconductivity. It is featured by low-energy spin excitations where exotic excitations such as fractional quantum numbers are easily induced by low energy. However, much of it is not understood and still hotly debated. One critical problem is the lack of a clean spin system that is free of chemical disorder and interference from conducting electrons. Meanwhile, the newly identified field-induced up-up-down (uud) order at ultralow temperatures in $Ba_3CoSb_2O_9$ [1-3], $AYbCh_2$ ($A$ = Na and Cs, $Ch$ = O, S, Se) [4-6], $Na_2BaNi(PO_4)_2$ [7], and $Na_3Fe(PO_4)_2$ [8] renewed attention on exotic phases in spin-$\frac{1}{2}$ triangular-lattice antiferromagnets. This field-induced uud order was theoretically predicted [9,10] but the experimental realization has been rare. Here, we report a unique zero-field Kagome up-up-down—down-down-up (uud-ddu) spin order in spin-$\frac{1}{2}$ Clinoatacamite, providing a prototypical model for comprehensively understanding pyrochlore quantum magnets in the picture of weakly-coupled Kagome layers as well as a new clue to the study of frustrated magnetism in Kagome antiferromagnetse.

The deformed pyrochlore-lattice Clinoatacamite $Cu_2(OH)_3Cl$, or more properly expressed as $Cu_4(OH)_6Cl_2$, reflecting its crystal structure, was viewed as the parent compound for the most researched spin liquid candidate Herbertsmithite $ZnCu_3(OH)_6Cl_2$ [11]. It is a geometrically frustrated spin-$\frac{1}{2}$ antiferromagnet with mysterious transitions from a seemingly complete long-range magnetic order to coexisting spin order/fluctuations [12]. The pyrochlore lattice of $Cu_4(OH)_6Cl_2$ consists of alternatively stacked triangular and Kagome layers, wherein selective replacement is enabled due to their different OH and Cl chemical environments for Cu. Thus, replacing the triangular-layer Cu with nonmagnetic Zn in $Cu_4(OH)_6Cl_2$ led to the birth of Herbertsmithite $ZnCu_3(OH)_6Cl_2$. One key issue in the spin dynamics of Herbertsmithite is the large number of defects in its Kagome layers due to an inevitable site mixing of Zn and Cu up to 15% [13,14]. Therefore, there is continuing attention to the intrinsic magnetism of Clinoatacamite for being a chemically clean parent compound of Herbertsmithite, which may provide a clue to solving some critical issues in this well-known spin-liquid candidate [15].

The spin behaviors in Clinoatacamite remained a mystery to date. Magnetic susceptibility and specific-heat measurements suggested antiferromagnetic transition with weak ferromagnetism below $T_{N2}$ ~ 6.4 K but only a tiny anomaly near $T_{N1}$ = 18.1 K [16]. The sensitive magnetic probe of muon spin spectroscopy ($\mu$SR) demonstrated a complete long-range order below $T_{N1}$, as can be seen in **SFig. 1**, which was a supplementary figure not shown previously due to space limitation [12], and then an unconventional transition from the complete long-range order to coexisting order/fluctuations below $T_{N2}$ [12]. For the low-temperature phase, neutron diffraction studies proposed several different magnetic structures varying from coexisting Néel order and valence-bond solid (VBS) state [17] to three-dimensional canted Néel orders [18,19]. A diffuse inelastic neutron scattering around 7 meV existed from the lowest temperature up to 30 K, suggesting inherent spin fluctuations [17,20]. The intermediate phase at $T_{N2} < T < T_{N1}$ is viewed as most puzzling. Although the $\mu$SR demonstrated a complete magnetic long-range order [12], only an almost unrecognizably small magnetic susceptibility change, which became visible in the differential of $\chi T$-$T$, and minor entropy release (~0.05$R$ln2) were observed at $T_{N1}$ [16]. All previous neutron diffraction experiments found no magnetic reflections for $T_{N2} < T < T_{N1}$ [17-19]. On this background, an unlikely hypothesis of stacking faults was even considered to explain this puzzling issue [15], despite the facts that similar compounds of Claringbullite $Cu_4(OH)_6ClF$ and Barlowite $Cu_4(OH)_6BrF$ also showed antiferromagnetic transitions at $T_{N1}$ = 17 K and 15 K, respectively [21, 22]. In the latter, the existence of possible site disorder in Claringbullite and Barlowite led to this unlikely hypothesis. On the other hand, a recent study on Ga substitution in $Ga_xCu_{4-x}(OH)_6Cl_2$ for $x$ = 0.19-0.80 contradicted this hypothesis [23]. In the present work, we have unambiguously solved this issue by uncovering a unique Kagome magnetic order in the intermediate phase of Clinoatacamite directly from neutron diffraction. Furthermore, magnetization measurements on a single crystalline Clinoatacamite have provided consistent picture of the magnetization in Clinoatacamite.

Polycrystalline Clinoatacamite $Cu_4(OD)_6Cl_2$ was synthesized by solution reaction of $CuCl_2$ and NaOD in a 1:1.2 mol ratio in $D_2O$ at 95 °C, followed by a hydrothermal treatment in $D_2O$ at 200 °C for one day. Small single crystals of Clinoatacamite $Cu_4(OH)_6Cl_2$ were successfully grown after many tries by heating the polycrystalline powder, which was synthesized as reported in ref. 14, in $CuCl_2$ solution at 340 °C in a hydrothermal furnace for 5-30 days followed by slow cooling. Clinoatacamite has stable polymorphous compounds atacamite and botallackite with features of geometric frustration [24]; therefore, careful characterization of the crystal structure is a necessity. An equivalent structure was confirmed for the polycrystalline $Cu_4(OD)_6Cl_2$ from neutron diffraction (**SFig. 2**) and single crystal $Cu_4(OH)_6Cl_2$ from synchrotron X-ray diffraction, as shown in **STables 1&2**, which was the same as the synthesized Clinoatacamite and the minerals [16, 25, 26]. The bond lengths and bonding angles between neighboring $Cu^{2+}$ ions are summarized in **STable 3**, giving a straightforward assessment of the super-exchange interactions. As illustrated in S**Fig. 3**, four $Cu^{2+}$ ions of site Cu1, Cu2, and two symmetric Cu3 form a deformed tetrahedron, in which any three Cu constitute a Kagome lattice plane. Of them, only the Cu1 and Cu3 ions in the (101) lattice plane are bonded via a single Cu-O-Cu bridge wherein the magnetic interactions are not frustrated.

As shown in Fig. 1, neutron diffraction patterns clearly demonstrated an antiferromagnetic long-range order, with distinct magnetic reflections indexed as $k$ = (-1/2 0 1/2), developed in the intermediate phase, consistent with the $\mu$SR experiment. As shown in detail in **STable 4**, for the magnetic propagation vector $k$ = (-1/2 0 1/2) in the $P2_1/n$ crystal space group, the possible magnetic structures were represented as $\Gamma_{mag}$ = $0\Gamma_1(A_g) + 3\Gamma_2 (A_u) + 0\Gamma_3 (B_g) + 3\Gamma_4$ ($B_u$), $3\Gamma_1 + 0\Gamma_2 + 3\Gamma_3 + 0\Gamma_4$, $3\Gamma_1 + 3\Gamma_2 + 3\Gamma_3 + 3\Gamma_4$, respectively, for the Cu1 (2$d$ position in STable 1), Cu2 (2$a$ position) and Cu3 (4$e$ position) sites, where $\Gamma_i$s are the irreducible representations [27]. Therefore, possible magnetic orders are allowed on sites Cu1-Cu3 with $\Gamma_2$ or $\Gamma_4$ in the Kagome plane (101) or on Cu2-Cu3 with $\Gamma_1$ or $\Gamma_3$ in the Kagome plane (10-1). The fitting results for the symmetry-allowed structures, as shown in **Fig. 2**, suggest that the $\Gamma_4$ with a contrastingly smallest $\chi^2$ = 0.5916 is the correct magnetic structure. This is in perfect consistency with an expectation from the crystal structure that only the Cu1 and Cu3 ions in the (101) lattice plane are bonded via a single Cu-O-Cu bridge (while all other pairs are frustrated with double ~ 90º bridges). The magnetic moments were $M_x$ = 0.107 (10) $\mu_B$, $M_y$ = 0.22 (2) $\mu_B$, $M_z$ = -0.11 (10) $\mu_B$, $M_{total}$ = 0.245(15) $\mu_B$ and $M_x$ = 0.05 (20) $\mu_B$, $M_y$ = 0.389 (3) $\mu_B$, $M_z$ = -0.01 (2) $\mu_B$, $M_{total}$ = 0.39(3) $\mu_B$, respectively, for the Cu1 and Cu3, indicating a noncoplanar spin texture. There are small uncancelled magnetic moments along the $a$-axis for the Cu1 spins. As shown in STable 3 and more visually seen in **Fig. 3**, the Cu1-Cu3 were bonded alternatively by O1 and O3 with slightly different bonding angles. The Cu1-Cu3 bonded by ∠Cu1-O1-Cu3 = 123.87(11)º are nearly antiparallel and those bonded by ∠Cu1-O3-Cu3 = 117.02(10)º are nearly parallel. This, again, agrees with a structural expectation from the super-exchange interactions.

Magnetization measurements performed on the single crystal of Clinoatacamite (7.3 mg), with the well-featured surfaces corresponding to lattice planes as indexed in **Fig. 4**, showed consistent changes at the transition at $T_{N1}$. Anisotropic susceptibility changes in $\chi T$ were observed along typical directions. The Cu3-Cu3 direction (the $b$-axis direction), the direction normal to the (011) lattice plane, and the Cu1-Cu2 (the $a$-axis) direction showed most clear changes at $T_{N1}$ =18.1 K, in good consistency to the revealed magnetic structure wherein the antiferromagnetic moments are predominantly along the $b$ and $a$-axis directions. Meanwhile, the Cu1-Cu3 and Cu2⊥Cu1-Cu3 directions, which are almost vertical to the antiferromagnetic spin directions, almost showed no change at $T_{N1}$. The $\chi$ kept rise below $T_{N1}$, reflecting the uncancelled nature of the Cu1 spins as revealed in Fig. 2. Susceptibilities at high temperatures suggested Curie-Weiss temperatures of $\theta_{CW}$ = -217(1) K, -161(1) K, -73(1) K, and -14(4) K, respectively, along directions Cu1-Cu3, the *b*-axis, the *a*-axis, and Cu2-Cu3, in agreement with the expectation from the Cu-O-Cu bonding.

The transition at $T_{N1}$ =18.1 K into the two-dimensional Kagome order is consistent with a previous Raman spectroscopy study wherein a broad continuum below 100 K reminiscent of low-dimensional quantum spin systems [28] was observed [29]. The present work revealed that the Cu2 spins in the triangular layer that links the (101) Kagome layers are free in the intermediated phase. Although the VBS model, as proposed by a previous neutron scattering study [17], is denied, the proposed concept of weakly coupled Kagome lattice has been verified. Previous specific heat measurements [30] and a numerical study of thermodynamics [31], which suggested a weak FM coupling ($J_{FM} = -0.1 J_{AFM}$) between the triangular and Kagome layers, also reinforced this concept. The Cu1-Cu2, as can be seen in Fig. 2 and SFig. 3, are bonded by two nearly 90º Cu-O-Cu bonds along the *a*-axis direction. Apparently, the Cu1-Cu2 FM coupling was the reason for the reported three-dimensional magnetic structure in the low-temperature phase wherein Cu1 and Cu2 were aligned in a ferromagnetic manner with the spin directions pointing to the *a*-axis [18]. With the Kagome order revealed, we can now explain the $\mu$SR results in a more profound insight view. The order below $T_{N1}$ =18.1 K seemed fcomplete in the $\mu$SR, as suggested by the oscillating spectra around the baseline of 1/3 $P_z(t)$ [12]. Now, this puzzle can be readily explained by paramagnetic Cu2 spins, which fluctuate too fast beyond the $\mu$SR time window. One can expect that the enhanced inter-Kagome-layer couplings would un-stabilize the two-dimensional Kagome order upon further cooling. Indeed, a nearly spin liquid state had been observed at $T \sim 5.7$ K, wherein the muon spins got almost completely depolarized [12] before the system finally reached a state of coexisting order and residual spin fluctuations.

In summary, a unique zero-field noncoplanar up-up-down—down-down-up Kagome spin order in spin-1/2 antiferromagnet Clinoatacamite, $Cu_4(OH)_6Cl_2$, which was known as the parent compound for the most researched spin liquid candidate Herbertsmithite $ZnCu_3(OH)_6Cl_2$, has been revealed. The new Kagome order in this chemically clean parent compound should contribute to the study of frustrated magnetism in the most-researched spin liquid candidate, Herbertsmithite, wherein inter-Kagome couplings are greatly weakened. The readily accessible temperature range for these quantum phases is also a merit for experimental studies. It is interesting to compare the present zero-field uud-ddu order to the field-induced uud order at ultralow temperatures in $Ba_3CoSb_2O_9$ [1-3], $A$YbCh$_2$ ($A$ = Na and Cs, $Ch$ = O, S, Se) [4-6], $Na_2BaNi(PO_4)_2$ [7], and $Na_3Fe(PO_4)_2$ [8]. We expect field-induced new phases in the present compound.

## ACKNOWLEDGMENTS

Synchrotron radiation X-ray diffraction experiments were carried out at SPring-8 with the approval of the Japan Synchrotron Radiation Research Institute (Proposal No. 2018B0078 and 2019A0159). The neutron diffraction experiments were performed at J-PARC under a user program (experiment Nos. 2014A0184, 2015A0236, 2017A0123, 2023B0128). The muon spin rotation/relaxation experiments at KEK-MSL were performed under a user program. We are grateful to Dr. S. Torii and Prof. T. Kamiyama at J-PARC for their technical support. We thank Prof. H. Kawamura for the fruitful discussion.

**Author contribution**: XGZ: conceptualization and writing, polycrystalline sample synthesis, magnetization, $\mu$SR and neutron diffraction experiments. MH: single crystal growth, final neutron diffraction experiment and magnetic structure refinement. IY: analysis of the $\mu$SR spectra. EN: single crystal synchrotron X-ray diffraction experiment and anylysis. TH: final neutron diffraction experiment. TY: single crystal growth for the synchrotron X-ray diffraction. CNX: assistance in hydrothermal reactions.

**Table and figure captions**

**Fig. 1** Neutron diffraction patterns at 7 K and 30 K for Clinoatacamite $Cu_4(OD)_6Cl_2$ obtained at the high-intensity neutron total diffractometer NOVA installed at BL21, J-PARC. The inset plot shows the diffraction patterns in a narrow $d$-range at the low-background SuperHRPD BL08 beamline. Distinct magnetic reflections indexed as $\boldsymbol{k}$ = (-1/2 0 1/2) were observed.

**Fig.2** The difference between the intensities obtained at 7 K and 30 K, and the fitting results for candidate magnetic structures $\Gamma_1$, $\Gamma_2$, $\Gamma_3$, and $\Gamma_4$ with $\chi^2$ = 1.5563, 1.9725, 2.3532, 0.5916, respectively. With the best fitting of $\Gamma_4$, magnetic moments of $M_x$ = 0.107 (10) $\mu_B$, $M_y$ = 0.22 (2) $\mu_B$, $M_z$ = -0.11 (10) $\mu_B$, $M_{total}$ = 0.245(15) $\mu_B$ and $M_x$ = 0.05 (20) $\mu_B$, $M_y$ = 0.389 (3) $\mu_B$, $M_z$ = -0.01 (2) $\mu_B$, $M_{total}$ = 0.39(3) $\mu_B$ were obtained for the Cu1 and Cu3, respectively. The masked areas were nuclear reflections remained after the subtraction due to the high background of the total diffractometer.

**Fig.3** The revealed alternative uud-ddu noncoplanar Kagome order in the lattice plane (101) in the intermediate phase in Clinoatacamite. Cu ions are represented in blue, and O ions in red. The Cu1 and Cu3 spins bonded via ∠Cu1-O1-Cu3 = 123.87(11)° are nearly antiparallel, and those bonded via ∠Cu1-O3-Cu3 = 117.02(10)° are nearly parallel. Meanwhile, the neighboring Cu3 spins bonded via ∠Cu3-O2-Cu3 = 116.91(9)° are always nearly antiparallel. These (101) Kagome planes are weakly coupled by the out-of-plane paramagnetic Cu2, and a more comprehensive view can be found in SFig. 3.

**Fig. 4** Magnetic susceptibilities measured on a single crystal of Clinoatacamite $Cu_4(OH)_6Cl_2$. Transition at $T_{N1}$ = 18.1 K with anisotropic changes in $\chi T - T$ can be clearly seen. Meanwhile, a further transition at $T_{N2}$ ~ 6 K is also witnessed. In the plot, the crystallographic directions are denoted referring to the labeled atoms in the crystal structure, wherein Cu1-Cu2 coincides with the $a$-axis direction, Cu3-Cu3 the $b$-axis, Cu2⊥Cu1-Cu3 implies the direction vertical to the Cu1-Cu3 line in the (011) lattice plane, and ⊥(011) the direction normal to the (011) lattice plane.

**STable 1**. Structure information for Cu$_4$(OD)$_6$Cl$_2$ at 30 K.

**STable 2**. Structure information for single crystal Clinoatacamite Cu$_4$(OH)$_6$Cl$_2$ at 100 K.

**STable 3**. Bond lengths and bond angles between neighboring Cu and O ions in Clinoatacamite Cu$_4$(OH)$_6$Cl$_2$ (at 100 K) for an easy-view assessment of the super-exchange interactions. The pairs of Cu1=Cu2 and Cu2=Cu3 are bonded via two nearly vertical Cu-O-Cu bridges, wherein frustrated magnetism can be expected. The underlined pairs of Cu1-Cu3 and Cu3-Cu3 are bonded via a single Cu-O-Cu bridge with larger bonding angles. Therefore, antiferromagnetic interactions are expected to be dominating.

**STable 4**. The Basis vectors for the intermediate phase with the magnetic propagation vector $\boldsymbol{k}$ = (-1/2 0 1/2) in the $P2_1/n$ crystal space group Cu$_4$(OD)$_6$Cl$_2$. The $x$, $y$, and $z$ represent the fractional coordinates ($x$ = 0.24228, $y$ = 0.23560, and $z$ = 0.25147).

**SFig. 1** $\mu$SR spectra showing the transition into a long-range order at $T_{N1} \sim$ 18.1 K in Clinoatacamite Cu$_4$(OH)$_6$Cl$_2$. The thick line is a fitted curve of the Kubo-Toyabe function combined with a 2-spin model function, *i.e.*, $P_z(t) = e^{-\Lambda t} [0.649\, \boldsymbol{G_Z^{KT}(t,\Delta)} + \boldsymbol{0.351 G_{2s}(t)}]$. The inset shows the temperature dependence of $\Lambda$ near $T_{N1}$, wherein the depolarization function was replaced by $\boldsymbol{P_Z(t) = P_Z(0)\left[\frac{1}{3}e^{-\Lambda t} + \frac{2}{3}e^{-\lambda t}\cos(\omega t + \phi)\right]}$ for $T < T_{N1}$. The observed muon spin polarization in the paramagnetic state close to $T_{N1}$ was previously expressed by $P_Z(t) = e^{-\Lambda t} G_Z^{KT}(t,\Delta)$, wherein the exponential function represents the relaxation due to electron spins of Cu$^{2+}$ and the Kubo-Toyabe function $\boldsymbol{G_Z^{KT}}$ the relaxation due to the nuclear fields produced by H and Cl nuclei (*12*). An exact model has been established with $\mu$SR spectra obtained at a long-time pulsed muon facility. The previously used Toyabe function $\boldsymbol{G_Z^{KT}}$ should be more precisely replaced with the sum of the $\boldsymbol{G_Z^{KT}}$ and a 2-spin model function $\boldsymbol{G_{2s}(t)}$ actually, 64.9% of the implanted muons were found to be best fitted by the Kubo-Toyabe function $\boldsymbol{G_Z^{KT}}$ with $\Delta$ = 0.39 (1) μs$^{-1}$. The remaining 35.1% were found to be stopped near OH$^-$, forming (OH)−$\mu^+$ bonding, as was often seen in materials containing hydrogen [32,33]. In the latter, its behavior was well described by a 2-spin model function $\boldsymbol{G_{2s}(t)} = \frac{1}{6} + \frac{1}{6}\cos(\omega t) + \frac{1}{3}\cos(\frac{1}{2}\omega t) + \frac{1}{3}\cos(\frac{3}{2}\omega t)$, where $\omega = \hbar\gamma_\mu\gamma_N/r^3$, $\gamma_\mu$ and $\gamma_N$ the gyromagnetic ratios of $\mu^+$ and H$^+$ nuclear spins, respectively. The distance between $\mu^+$ and $H^+$ in the present compound was fitted to be $r = 1.47(1)$ Å.

**SFig. 2** Neutron diffraction pattern of Clinoatacamite $Cu_4(OD)_6Cl_2$ at 30 K obtained from the BS and LA banks at SuperHRPD, J-PARC. The solid lines are fitted by Rietveld refinement with parameters in STable 2.

**SFig. 3** Crystal structure of Clinoatacamite $Cu_4(OH)_6Cl_2$. The magnetic ions of $Cu^{2+}$ presented in blue are denoted with their site numbers 1,2,3. Only the Cu1 and Cu3 ions in the (101) lattice plane are bonded via a single Cu-O-Cu bridge wherein the magnetic interactions are not frustrated. The Cu1 and Cu3 ions form a Kagome lattice in the (101) lattice plane.

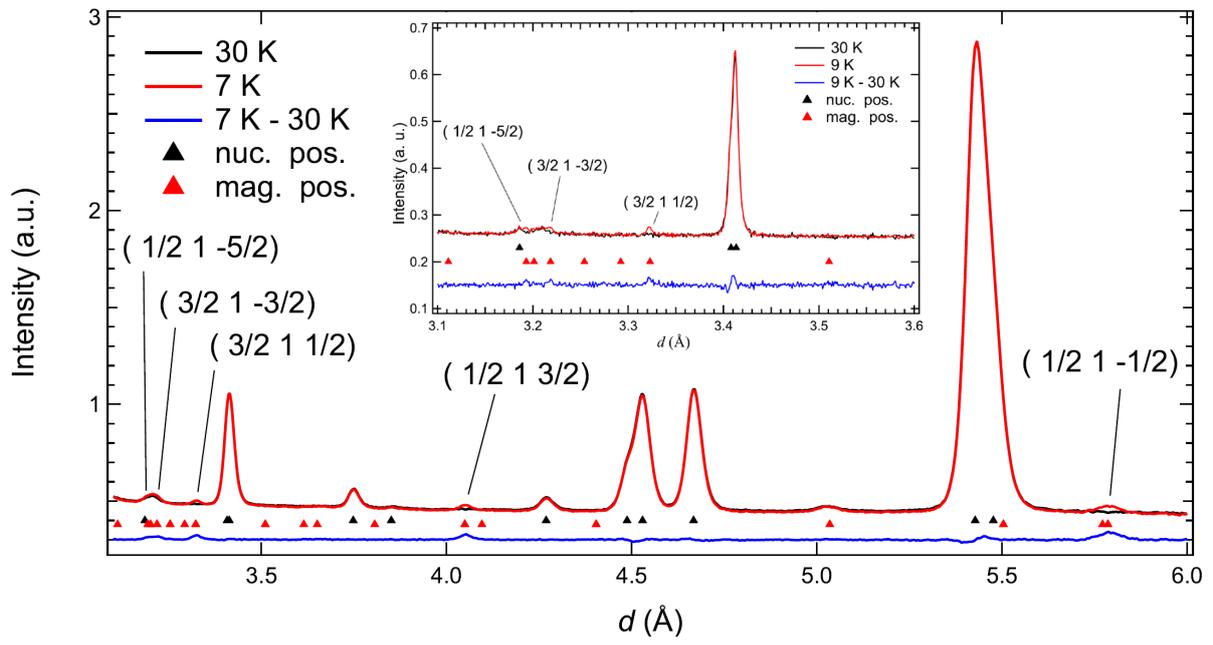

Fig. 1

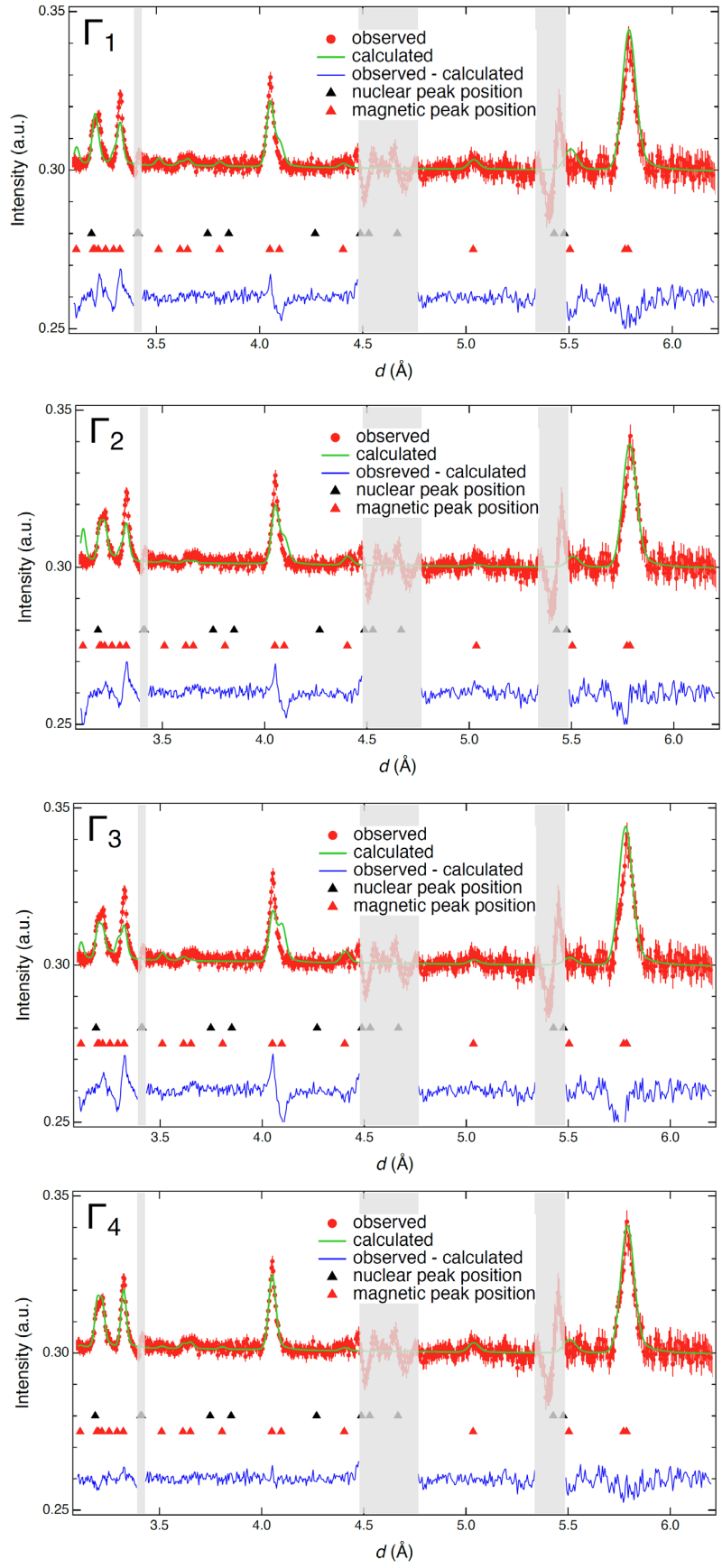

Fig. 2

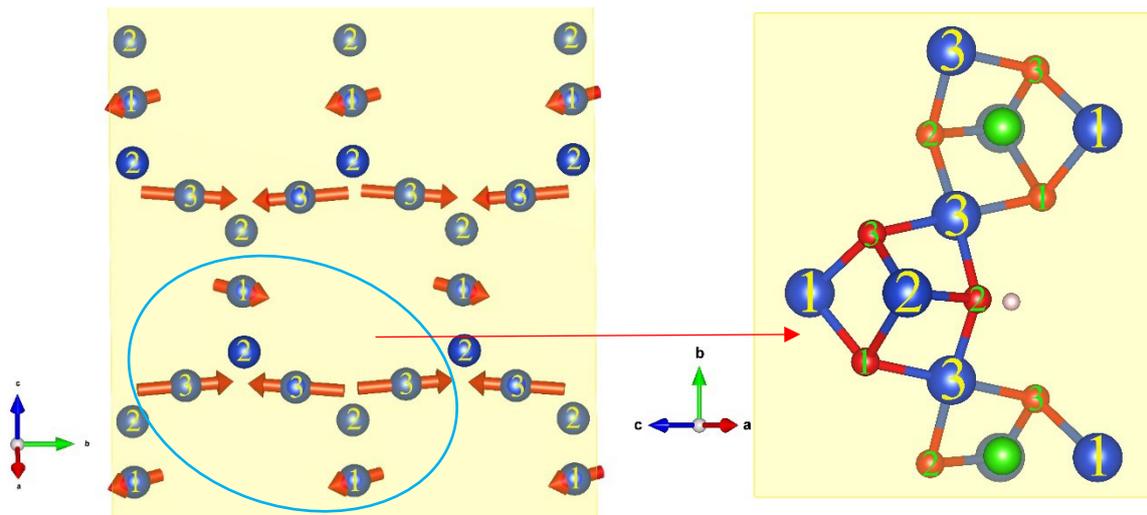

Fig. 3

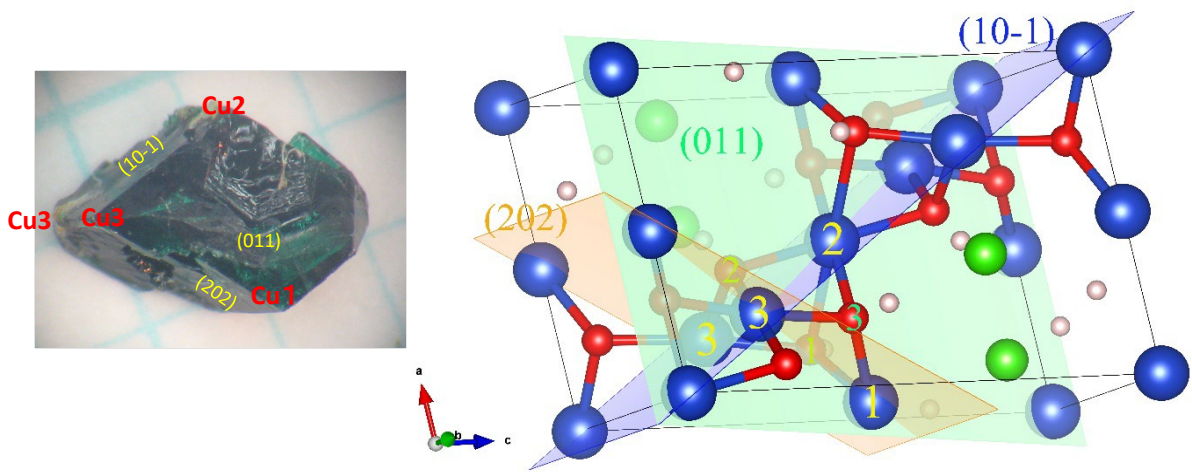
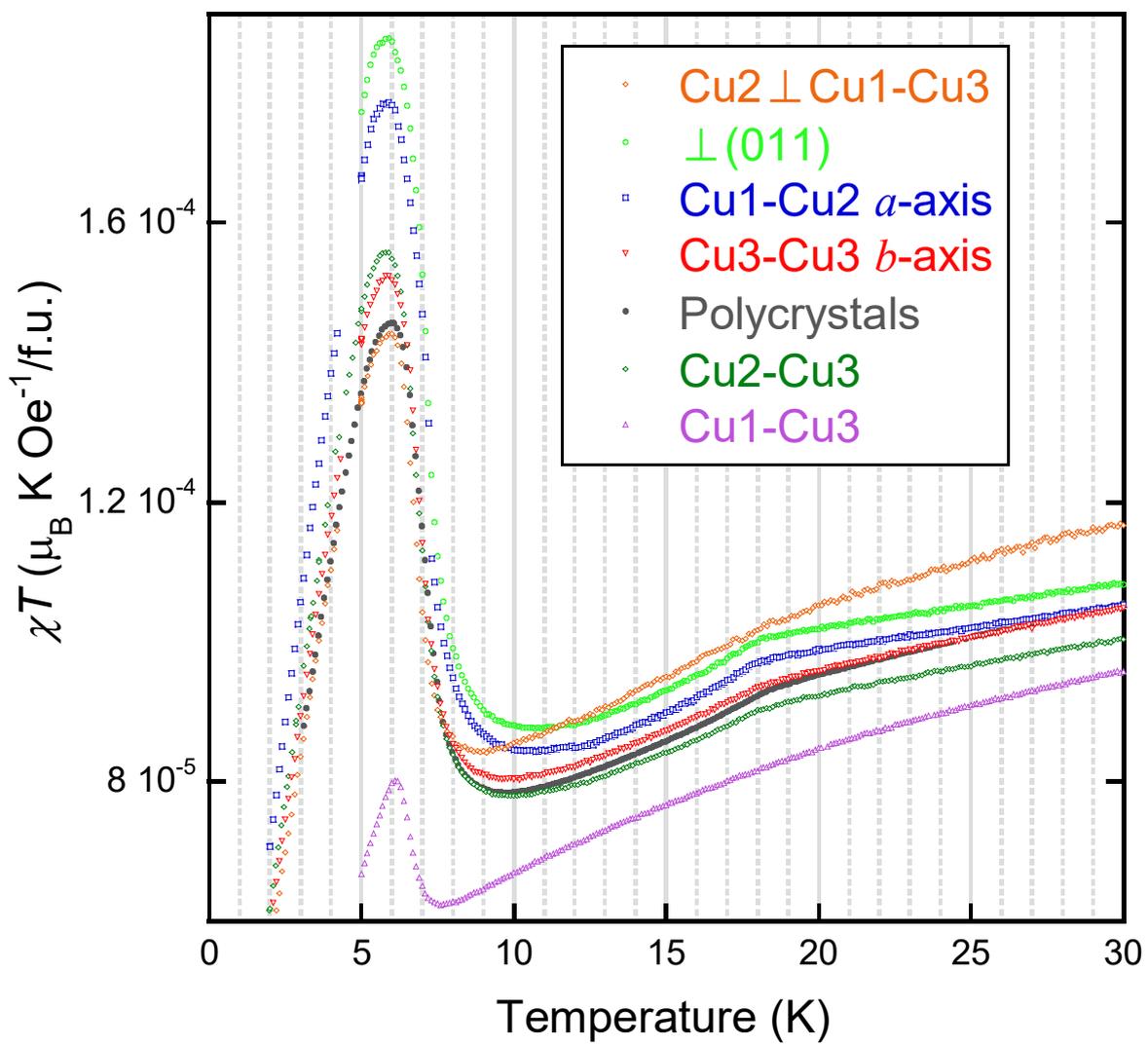

Fig. 4

**STable 1.** Structure information of Clinoatacamite $Cu_4(OD)_6Cl_2$ at 30 K.

| | |
|---|---|
| Chemical Formula | $Cu_4(OD)_6Cl$ |
| Cell Setting | Monoclinic |
| Space group | $P2_1/n$ (14-2) |
| $a$ (Å) | 6.152535 (3) |
| $b$ (Å) | 6.814393 (4) |
| $c$ (Å) | 9.109159 (4) |
| $\beta$ (deg.) | 99.84173 (5)° |
| Fitting range (Å) | 0.658~3.777 (BS) |
| | 2.45~11.16 (LA) |
| $R_{wp}$ (%) /$R_e$ (%) | 4.184/0.512 |
| $R_{wp}$ (%) /$R_e$ (%) | 4.422/0.518 (BS) |
| | 1.274/0.434 (LA) |
| $R_B$ (%) | 2.925 (BS) |
| | 1.896 (LA) |
| $R_F$ (%) | 2.386 (BS) |
| | 1.906 (LA) |
| $\chi^2$ | 66.70 |

| | Site | Sym | $x$ | $y$ | $z$ | $B$ (Å$^2$) | $g$ |
|---|---|---|---|---|---|---|---|
| Cu1 | 2d | -1 | 0.5 | 0 | 1 | 0.057(2) | 1 |
| Cu2 | 2a | -1 | 0.5 | 0.5 | 0.5 | 0.057(2) | 1 |
| Cu3 | 4e | 1 | 0.74228(4) | 0.26440(4) | 0.75147(3) | 0.057(2) | 1 |
| Cl | 4e | 1 | 1.11228(2) | 0.49549(3) | 0.807288(14) | 0.123(3) | 1 |
| O1 | 4e | 1 | 0.74174(5) | 0.17909(4) | 0.95918(4) | 0.113(3) | 1 |
| O2 | 4e | 1 | 0.42224(4) | 0.51915(4) | 0.27645(4) | 0.113(3) | 1 |
| O3 | 4e | 1 | 0.80989(4) | 0.29050(5) | 0.54900(3) | 0.113(3) | 1 |
| D1 | 4e | 1 | 0.55516(4) | 0.52177(6) | 0.23027(3) | | 0.9004(3) |
| H1 | 4e | 1 | 0.55516(4) | 0.52177(6) | 0.23027(3) | | 0.0996(3) |
| D2 | 4e | 1 | 0.76285(5) | 0.29225(4) | 1.02851(5) | | 0.90004(3) |
| H2 | 4e | 1 | 0.76285(5) | 0.29225(4) | 1.02851(5) | | 0.0996(3) |
| D3 | 4e | 1 | 0.77837(6) | 0.18483(5) | 0.47657(4) | | 0.90004(3) |
| H3 | 4e | 1 | 0.77837(6) | 0.18483(5) | 0.47657(4) | | 0.0996(3) |

| | $\beta_{11}$ | $\beta_{22}$ | $\beta_{33}$ | $\beta_{12}$ | $\beta_{13}$ | $\beta_{23}$ |
|---|---|---|---|---|---|---|
| D1 | 0.189(3) | 0.147(3) | 0.0873(15) | 0.055(3) | 0.0883(15) | -0.017(2) |
| H1 | 0.189(3) | 0.147(3) | 0.0873(15) | 0.055(3) | 0.0883(15) | -0.017(2) |
| D2 | 0.222(3) | 0.070(3) | 0.153(2) | -0.019(3) | 0.043(2) | -0.0547(19) |
| H2 | 0.222(3) | 0.070(3) | 0.153(2) | -0.019(3) | 0.043(2) | -0.0547(19) |
| D3 | 0.338(4) | 0.130(3) | 0.050(2) | -0.043(3) | 0.029(2) | -0.0080(19) |
| H3 | 0.338(4) | 0.130(3) | 0.050(2) | -0.043(3) | 0.029(2) | -0.0080(19) |

**STable 2**. Structure information for single crystal Clinoatacamite $Cu_4(OH)_6Cl_2$ at 100 K.

| | | Chemical Formula | | | $Cu_4(OH)_6Cl_2$ | | |
|---|---|---|---|---|---|---|---|
| | | $Z$ | | | 4 | | |
| | | $F(000)$ | | | 408 | | |
| | | Density | | | 3.785 Mg m$^{-3}$ | | |
| | | Cell Setting | | | Monoclinic | | |
| | | Space group | | | $P2_1/n$, No. 14 | | |
| | | $a$ (Å) | | | 6.1459 (4) | | |
| | | $b$ (Å) | | | 6.8034 (4) | | |
| | | $c$ (Å) | | | 9.0960 (5) | | |
| | | $\beta$ | | | 99.854 (7)° | | |
| | Site | Sym | $x$ | $y$ | $z$ | $g$ | $U$ |
| Cu1 | 2$d$ | -1 | 0.50000 | 0.00000 | 1.000 | 1.000 | 0.003 |
| Cu2 | 2$a$ | -1 | 0.50000 | 0.50000 | 0.50000 | 1.000 | 0.003 |
| Cu3 | 4$e$ | 1 | 0.74108 | 0.26507 | 0.75158 | 1.000 | 0.003 |
| Cl | 4$e$ | 1 | 1.11210 | 0.49546 | 0.80747 | 1.000 | 0.007 |
| O1 | 4$e$ | 1 | 0.69050 | 0.20958 | 0.95140 | 1.000 | 0.006 |
| O2 | 4$e$ | 1 | 0.42220 | 0.48101 | 0.27665 | 1.000 | 0.004 |
| O3 | 4$e$ | 1 | 0.75790 | 0.32132 | 0.53928 | 1.000 | 0.004 |
| H1 | 4$e$ | 1 | 0.52000 | 0.47300 | 0.23700 | 1.000 | 0.011 |
| H2 | 4$e$ | 1 | 0.70300 | 0.33700 | 1.01100 | 1.000 | 0.012 |
| H3 | 4$e$ | 1 | 0.72900 | 0.20200 | 0.48500 | 1.000 | 0.011 |

**STable 3**. Bond lengths and bond angles between neighbouring Cu and O ions in Clinoatacamite $Cu_4(OH)_6Cl_2$ (at 100 K) for an easy-view assessment of the super-exchange interactions. The pairs of Cu1=Cu2 and Cu2=Cu3 are bonded via two nearly vertical Cu-O-Cu bridges, wherein frustrated magnetism can be expected. The underlined pairs of Cu1-Cu3 and Cu3-Cu3 are bonded via a single Cu-O-Cu bridge with larger bonding angles. Therefore, antiferromagnetic interactions are expected to be dominating.

Cu1 = O1O3 = Cu2
$d_{Cu1-O3}$ = 2.0003(13) Å, $d_{O3-Cu2}$ = 1.9809(13) Å, $\angle_{Cu1-O3-Cu2}$ = 101.04(9)°
$d_{Cu1-O1}$ = 1.9433(15) Å, $d_{O1-Cu2}$ = 2.3577(16) Å, $\angle_{Cu1-O1-Cu2}$ = 90.67(7)°

Cu2 = O1O2 = Cu3, Cu2 = O2O3 = Cu3

$d_{Cu2-O3}$ = 1.9809(13) Å, $d_{O3-Cu3}$ = 1.9886(13) Å, $\angle_{Cu2-O3-Cu3}$ = 96.91(9)°
$d_{Cu2-O2}$ = 2.0100(16) Å, $d_{O2-Cu3}$ = 1.9923(13) Å, $\angle_{Cu2-O2-Cu3}$ = 95.85(9)°
$d_{Cu2-O1}$ = 2.3577(16) Å, $d_{O1-Cu3}$ = 1.9329(13) Å, $\angle_{Cu2-O1-Cu3}$ = 92.29(10)°
$d_{Cu2-O2}$ = 2.0100(16) Å, $d_{O2-Cu3}$ = 2.0018(13) Å, $\angle_{Cu2-O2-Cu3}$ = 101.56(19)°

<u>Cu1 – O1 – Cu3, Cu1 – O3 – Cu3</u>
<u>$d_{Cu1-O3}$ = 2.0003(13) Å, $d_{O3-Cu3}$ = 1.9886(13) Å, $\angle_{Cu1-O3-Cu3}$ = 117.02(10)°</u>
<u>$d_{Cu1-O1}$ = 1.9433(15) Å, $d_{O1-Cu3}$ = 1.9329(13) Å, $\angle_{Cu1-O1-Cu3}$ = 123.87(11)°</u>

<u>Cu3 – O2 – Cu3</u>
<u>$d_{Cu3-O2}$ = 1.9923(13) Å, $d_{O2-Cu3}$ = 2.0018(13) Å, $\angle_{Cu3-O2-Cu3}$ = 116.91(9)°</u>

**STable 4**. The Basis vectors for the intermediate phase with the magnetic propagation vector $k$ = (-1/2 0 1/2) in the $P2_1/n$ crystal space group $Cu_4(OD)_6Cl_2$. The $x$, $y$, and $z$ represent the fractional coordinates ($x$ = 0.24228, $y$ = 0.23560 and $z$ = 0.25147).

| IR | BV | site Cu1 $[\frac{1}{2}, 0, 0]$ | Cu1 $[0, \frac{1}{2}, \frac{1}{2}]$ | site Cu2 $[0, 0, 0]$ | Cu2 $[\frac{1}{2}, \frac{1}{2}, \frac{1}{2}]$ |
|---|---|---|---|---|---|
| $\Gamma_1$ | $\Psi_1$ | - | - | (1, 0, 0) | (-1, 0, 0) |
| | $\Psi_2$ | - | - | (0, 1, 0) | (0, 1, 0) |
| | $\Psi_3$ | - | - | (0, 0, 1) | (0, 0, -1) |
| $\Gamma_2$ | $\Psi_4$ | (1, 0, 0) | (-1, 0, 0) | - | - |
| | $\Psi_5$ | (0, 1, 0) | (0, 1, 0) | - | - |
| | $\Psi_6$ | (0, 0, 1) | (0, 0, -1) | - | - |
| $\Gamma_3$ | $\Psi_7$ | - | - | (1, 0, 0) | (1, 0, 0) |
| | $\Psi_8$ | - | - | (0, 1, 0) | (0, -1, 0) |
| | $\Psi_9$ | - | - | (0, 0, 1) | (0, 0, 1) |
| $\Gamma_4$ | $\Psi_{10}$ | (1, 0, 0) | (1, 0, 0) | - | - |
| | $\Psi_{11}$ | (0, 1, 0) | (0, -1, 0) | - | - |
| | $\Psi_{12}$ | (0, 0, 1) | (0, 0, 1) | - | - |

| IR | BV | site Cu3 $[x, y, z]$ | $[1-x, 1-y, 1-z]$ | $[x+\frac{1}{2}, -y+\frac{1}{2}, z+\frac{1}{2}]$ | $[-x+\frac{1}{2}, y+\frac{1}{2}, -z+\frac{1}{2}]$ |
|---|---|---|---|---|---|
| $\Gamma_1$ | $\Psi_{13}$ | (1, 0, 0) | (-1, 0, 0) | (1, 0, 0) | (-1, 0, 0) |
| | $\Psi_{14}$ | (0, 1, 0) | (0, 1, 0) | (0, 1, 0) | (0, 1, 0) |
| | $\Psi_{15}$ | (0, 0, 1) | (0, 0, -1) | (0, 0, 1) | (0, 0, -1) |
| $\Gamma_2$ | $\Psi_{16}$ | (1, 0, 0) | (-1, 0, 0) | (-1, 0, 0) | (1, 0, 0) |
| | $\Psi_{17}$ | (0, 1, 0) | (0, 1, 0) | (0, -1, 0) | (0, -1, 0) |
| | $\Psi_{18}$ | (0, 0, 1) | (0, 0, -1) | (0, 0, -1) | (0, 0, 1) |
| $\Gamma_3$ | $\Psi_{19}$ | (1, 0, 0) | (1, 0, 0) | (1, 0, 0) | (1, 0, 0) |
| | $\Psi_{20}$ | (0, 1, 0) | (0, -1, 0) | (0, 1, 0) | (0, -1, 0) |
| | $\Psi_{21}$ | (0, 0, 1) | (0, 0, 1) | (0, 0, 1) | (0, 0, 1) |
| $\Gamma_4$ | $\Psi_{22}$ | (1, 0, 0) | (1, 0, 0) | (-1, 0, 0) | (-1, 0, 0) |
| | $\Psi_{23}$ | (0, 1, 0) | (0, -1, 0) | (0, -1, 0) | (0, 1, 0) |
| | $\Psi_{24}$ | (0, 0, 1) | (0, 0, 1) | (0, 0, -1) | (0, 0, -1) |

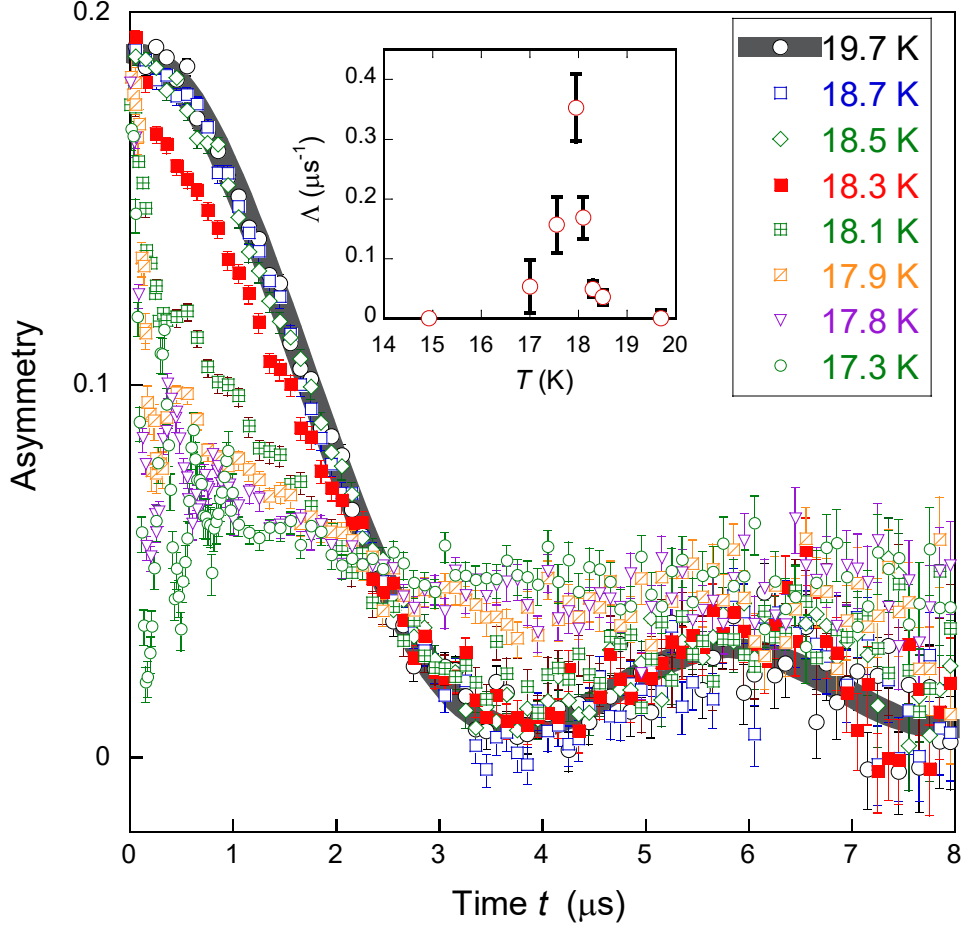

**SFig. 1** $\mu$SR spectra showing the transition into long-range order at $T_{N1} \sim 18.1$ K in Clinoatacamite $Cu_4(OH)_6Cl_2$. The thick line in is a fitted curve of the Kubo-Toyabe function combined with a 2-spin model function, *i.e.*, $P_Z(t) = e^{-\Lambda t}[0.649\, \boldsymbol{G_Z^{KT}(t,\Delta)} + \boldsymbol{0.351 G_{2s}(t)}]$. The inset shows the temperature dependence of $\Lambda$ near $T_{N1}$, wherein the depolarization function was replaced by $\boldsymbol{P_Z(t) = P_Z(0)\left[\frac{1}{3}e^{-\Lambda t} + \frac{2}{3}e^{-\lambda t}\cos(\omega t + \phi)\right]}$ for $T < T_{N1}$. The observed muon spin polarization in the paramagnetic state close to $T_{N1}$ was previously expressed by $P_Z(t) = e^{-\Lambda t} G_Z^{KT}(t,\Delta)$, wherein the exponential function represents the relaxation due to electron spins of $Cu^{2+}$ and the Kubo-Toyabe function $G_Z^{KT}$ the relaxation due to the nuclear fields produced by H and Cl nuclei (*12*). An exact model has been established with $\mu$SR spectra obtained using a long-time pulsed muon facility. The previously used Toyabe function $\boldsymbol{G_Z^{KT}}$ should be more precisely replaced with the sum of the $\boldsymbol{G_Z^{KT}}$ and a 2-spin model function $\boldsymbol{G_{2s}(t)}$ actually, 64.9% of the implanted muons were found to be best fitted by the Kubo-Toyabe function $\boldsymbol{G_Z^{KT}}$ with $\Delta = 0.39\,(1)$ $\mu$s$^{-1}$. The remaining 35.1% were found to be stopped near OH$^-$, forming (OH)--$\mu^+$ bonding, as was often seen in materials containing hydrogen [32,33]. In the latter, its behavior was well described by a 2-spin model function $\boldsymbol{G_{2s}(t) = \frac{1}{6} + \frac{1}{6}\cos(\omega t) + \frac{1}{3}\cos(\frac{1}{2}\omega t) + \frac{1}{3}\cos(\frac{3}{2}\omega t)}$, where $\omega = \hbar\gamma_\mu\gamma_N/r^3$, $\gamma_\mu$ and $\gamma_N$ the gyromagnetic ratios of $\mu^+$ and H$^+$ nuclear spins, respectively. The distance between $\mu^+$ and H$^+$ was fitted to be $r = 1.47(1)$ Å.

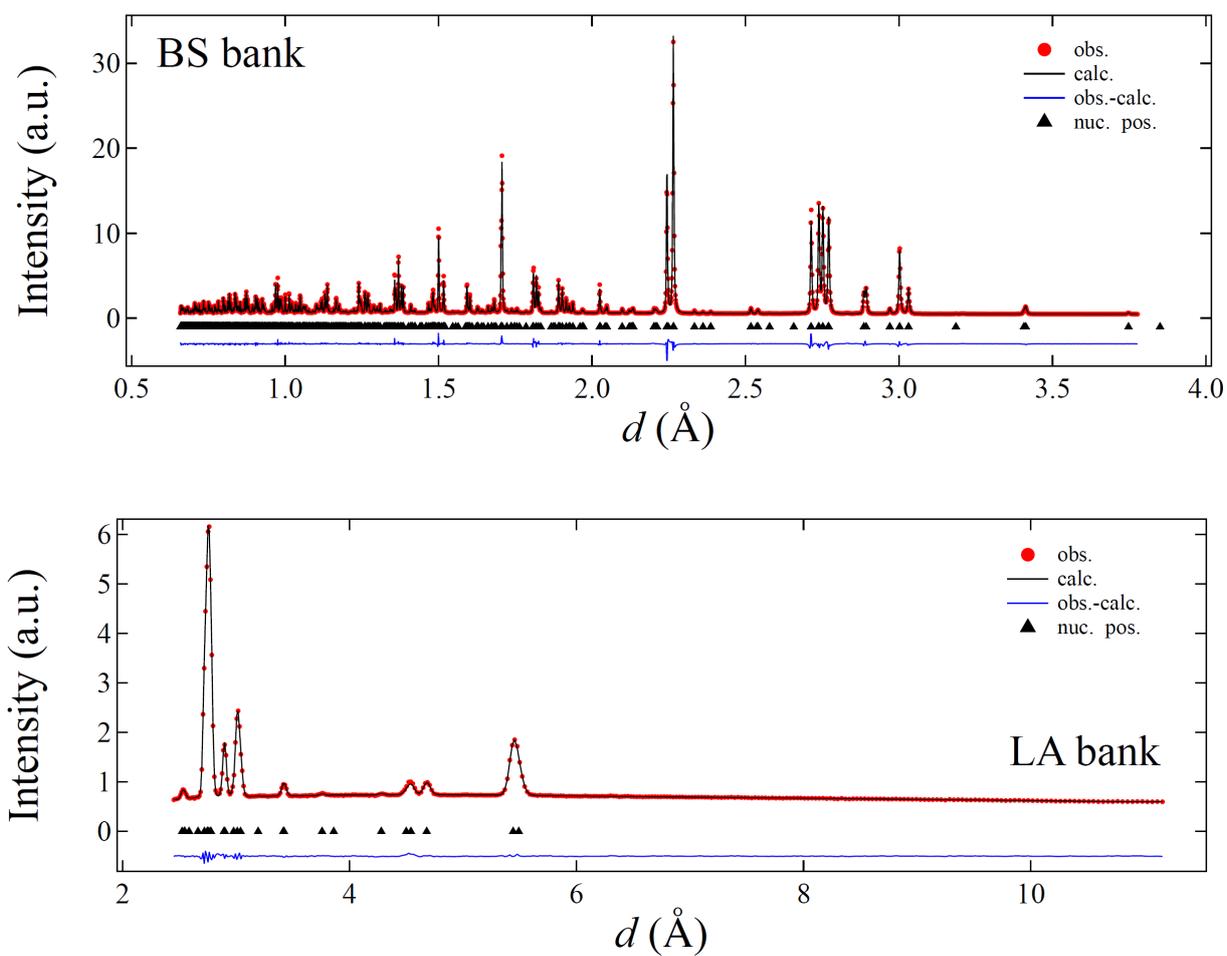

**SFig. 2** Neutron diffraction pattern of Clinoatacamite $Cu_4(OD)_6Cl_2$ at 30 K obtained from the BS and LA banks at SuperHRPD, J-PARC. The solid lines are fitted by Rietveld refinement with parameters in STable 2.

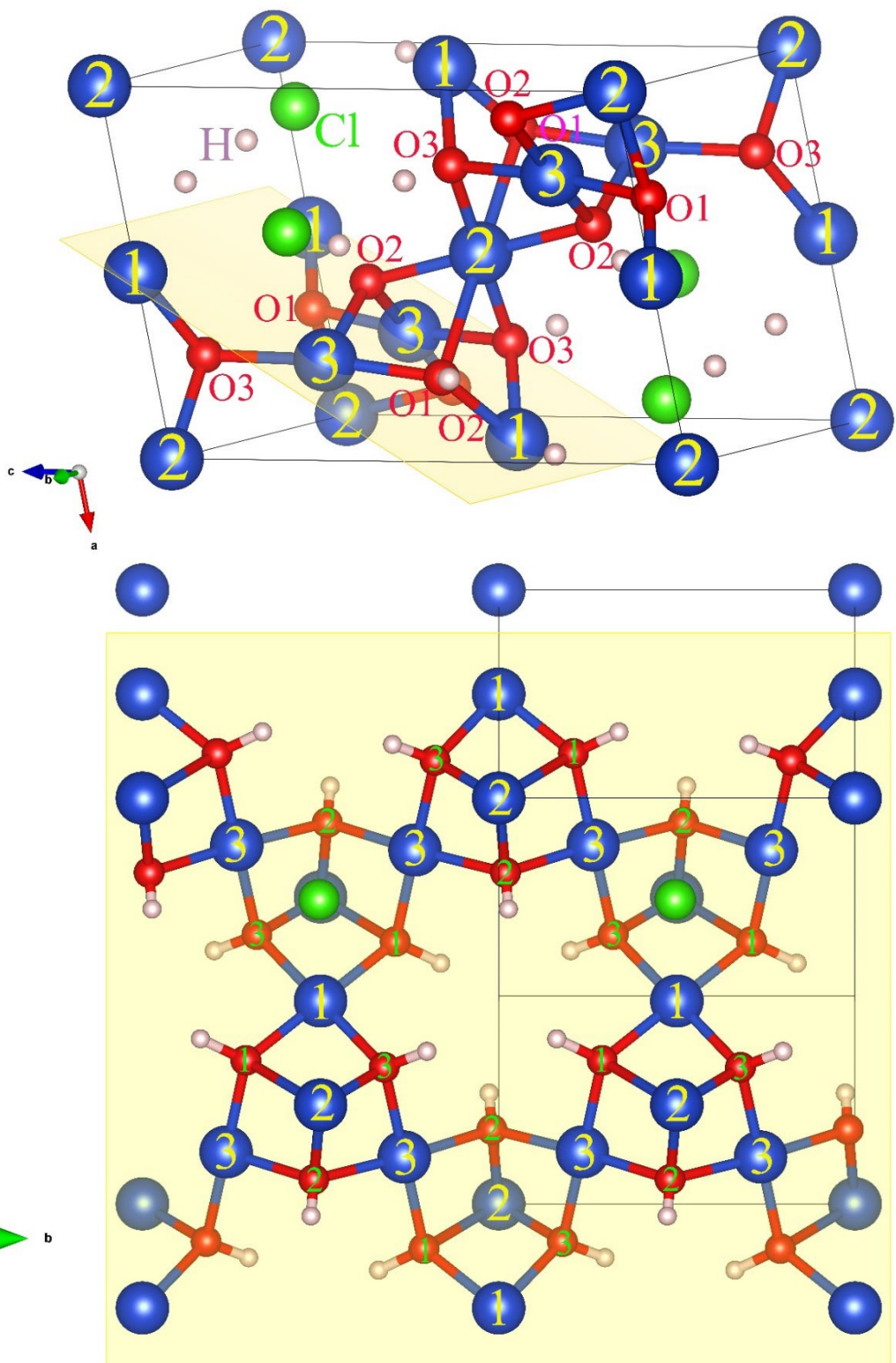

**SFig. 3** The upper panel: crystal structure of Clinoatacamite $Cu_4(OH)_6Cl_2$. The magnetic ions of $Cu^{2+}$ are presented in blue color labeled with their site numbers 1, 2, 3. Only the Cu1 and Cu3 ions in the (101) lattice plane are bonded via a single Cu-O-Cu bridge wherein antiferromagnetic interactions are expected to be dominating. The Cu1 and Cu3 ions form a Kagome lattice in the (101) lattice plane.

7